\documentclass[conference, a4paper]{IEEEtran}
\IEEEoverridecommandlockouts
\usepackage{svg}
\usepackage{multirow}
\usepackage{amsmath,amssymb,amsfonts,mathtools}
\usepackage{placeins}
\usepackage{algorithm,algpseudocode}
\usepackage{orcidlink} 
\usepackage[capitalise, noabbrev]{cleveref}
\usepackage[nolist]{acronym}

\usepackage{siunitx}
\usepackage{pifont}
\usepackage{subcaption}
\usepackage{graphicx}
\algrenewcommand\algorithmiccomment[1]{\hfill /*#1*/}

\def\BibTeX{{\rm B\kern-.05em{\sc i\kern-.025em b}\kern-.08em
    T\kern-.1667em\lower.7ex\hbox{E}\kern-.125emX}}

\bstctlcite{IEEEexample:BSTcontrol}

\makeatletter
\newcounter{problem}
\newcounter{tmp}
\newenvironment{problem}[1][htb]{%
    \renewcommand{\ALG@name}{Optimization Problem}
    \setcounter{tmp}{\thealgorithm}
    \setcounter{algorithm}{\theproblem}
   \begin{algorithm}[#1]%
  }{\end{algorithm}
  \setcounter{problem}{\thealgorithm}
  \setcounter{algorithm}{\thetmp}}
\makeatother

\usepackage[all]{background}
\usepackage{stackengine}
\setstackEOL{\\}
\setstackgap{L}{\normalbaselineskip}
\SetBgContents{\color{gray}{\tiny \Longstack{PREPRINT - Accepted at Design, Automation, and Test in Europe (DATE) Conference 2024}}}
\SetBgPosition{3.5cm,1cm}
\SetBgOpacity{1.0}
\SetBgAngle{0}
\SetBgScale{1.8}

\begin{document}
\bstctlcite{IEEEexample:BSTcontrol}

\begin{acronym}
    \acro{rram}[RRAM]{resistive random access memory}
    \acro{cim}[CIM]{computing-in-memory}
    \acro{mvm}[MVM]{matrix-vector multiplication}
    \acro{cnn}[CNN]{convolutional neural network}
    \acro{nn}[NN]{neural network}
    \acro{bfs}[BFS]{breadth-first search}
    \acro{ofm}[OFM]{output feature map}
    \acro{ifm}[IFM]{input feature map}
    \acro{bn}[BN]{batch normalization}
    \acro{gemm}[GEMM]{general matrix multiply}
    \acro{ilp}[ILP]{integer linear programming}
    \acro{minlp}[MINLP]{mixed-integer nonlinear programming}
    \acro{gpu}[GPU]{graphics processing unit}
    \acro{cpu}[CPU]{central processing unit}
    \acro{tpu}[TPU]{tensor processing unit}
    \acro{ml}[ML]{machine learning}
    \acro{asap}[ASAP]{as soon as possible}
    \acro{ilp}[ILP]{integer linear program}
    \acro{gpeu}[GPEU]{general purpose execution unit}
    \acro{sota}[SOTA]{state-of-the-art}
    \acro{noc}[NoC]{network on chip}
    \acro{pe}[PE]{processing element}
\end{acronym}

\title{
CLSA-CIM: A \underline{C}ross-\underline{L}ayer \underline{S}cheduling \underline{A}pproach for \underline{C}omputing-\underline{i}n-\underline{M}emory Architectures
}

\def\finalpaper{1}

\if\finalpaper1
{
    \author{
    \IEEEauthorblockN{Rebecca Pelke\orcidlink{0000-0001-5156-7072}, Jose Cubero-Cascante\orcidlink{0000-0001-9575-0856}, Nils Bosbach\orcidlink{0000-0002-2284-949X}, Felix Staudigl\orcidlink{0000-0001-9673-3070},
        Rainer Leupers\orcidlink{0000-0002-6735-3033}, Jan Moritz Joseph\orcidlink{0000-0001-8669-1225}}
    \IEEEauthorblockA{\textit{Institute for Communication Technologies and Embedded Systems}\\
        \textit{RWTH Aachen University, Germany}\\
        \{pelke, cubero, bosbach, staudigl, leupers, joseph\}@ice.rwth-aachen.de}
    \thanks{This work was funded by the Federal Ministry of Education and Research (BMBF, Germany) in the project NeuroSys (Project Nos. 03ZU1106CA).}
    \vspace{-1cm}
    }
}
\else
    \author{
      \IEEEauthorblockN{Authors are removed for submission version}
      \\
      \\
      \IEEEauthorblockA{Affiliations are removed for submission version}
      \vspace{-1cm}
      \\
    }
\fi

\maketitle

\begin{abstract}
The demand for efficient \ac{ml} accelerators is growing rapidly, driving the development of novel computing concepts such as \ac{rram}-based tiled \ac{cim} architectures. \ac{cim} allows to compute within the memory unit, resulting in faster data processing and reduced power consumption.
Efficient compiler algorithms are essential to exploit the potential of tiled \ac{cim} architectures. While conventional \acs{ml} compilers focus on code generation for CPUs, GPUs, and other von Neumann architectures, adaptations are needed to cover \ac{cim} architectures.
Cross-layer scheduling is a promising approach, as it enhances the utilization of \ac{cim} cores, thereby accelerating computations.
Although similar concepts are implicitly used in previous work, there is a lack of clear and quantifiable algorithmic definitions for cross-layer scheduling for tiled \ac{cim} architectures.

To close this gap, we present CLSA-CIM, a cross-layer scheduling algorithm for tiled \ac{cim} architectures.
We integrate CLSA-CIM with existing weight-mapping strategies and compare performance against \ac{sota} scheduling algorithms.
CLSA-CIM improves the utilization by up to \SI{17.9}{\times}, resulting in an overall speedup increase of up to \SI{29.2}{\times} compared to \ac{sota}.
\end{abstract}


\vspace{-0.2cm}
\section{Introduction}

\acresetall 

The increasing demand for efficient computation of data-intensive \ac{ml} applications has led to specialized architectures such as \acp{gpu} and \acp{tpu}. However, a major performance limitation is the data movement between main memory and compute units, known as the von Neumann bottleneck~\cite{zou2021breaking}. Novel \ac{cim} technologies, such as \ac{rram}, tackle this bottleneck by unifying memory and computation unit~\cite{chang2017memcomputing}.
These designs outperform their CMOS-based counterparts in memory capacity, device density, and power consumption~\cite{vetter2015opportunities}.

\begin{figure}[bt]
  \centering
  \includegraphics[width=.9\linewidth, keepaspectratio]{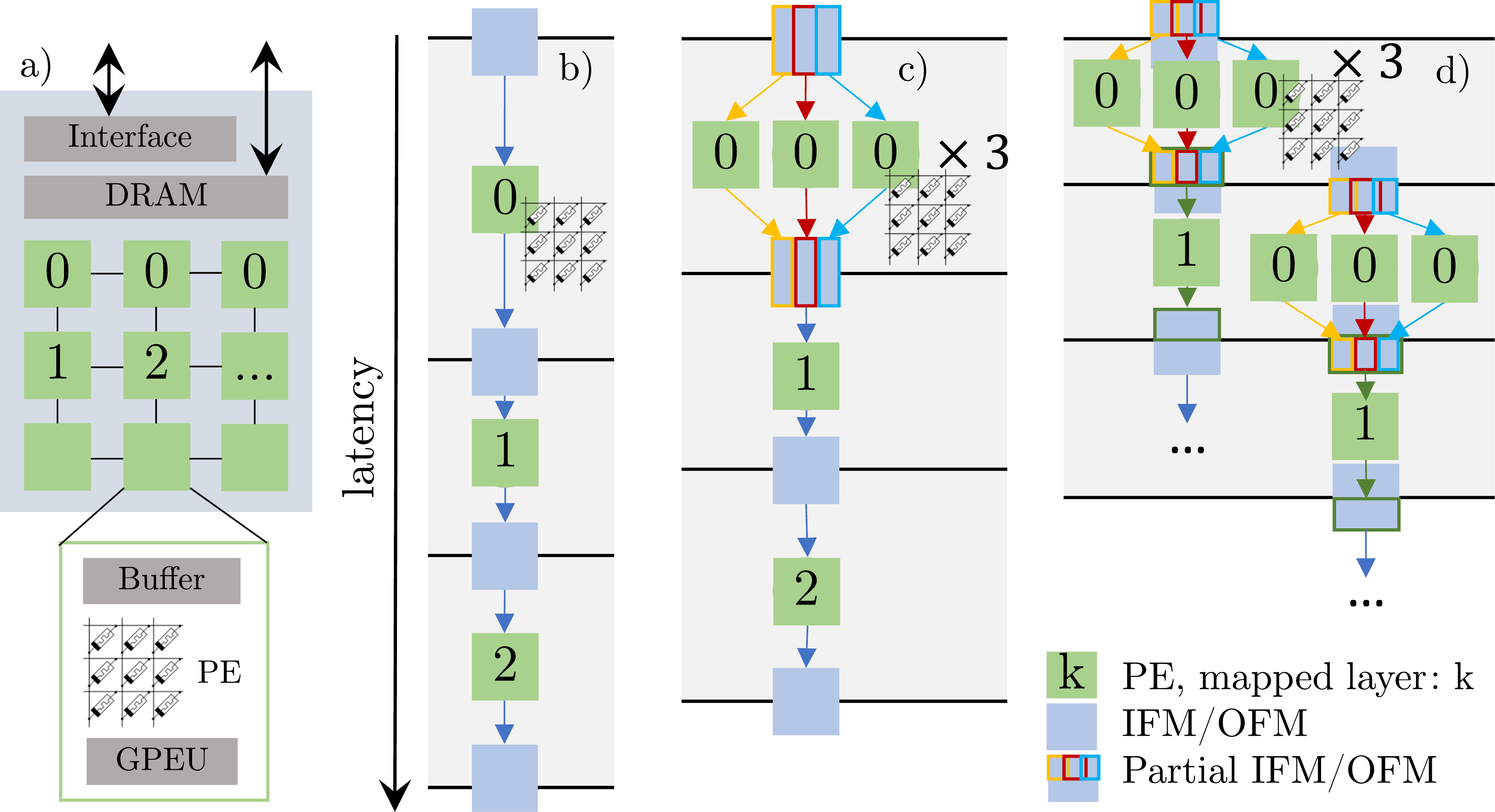}
  \caption{\acs{nn} inference on (a) tiled \acs{cim} architectures: (b) Layer-by-layer scheduling, (c) weight duplication mapping, and (d) cross-layer scheduling}
  \label{fig:overview}
  \vspace{-0.5cm}
\end{figure}

In recent years, several \ac{cim} architectures have been introduced~\cite{NeuRRAM,ankit2019puma,shafiee2016isaac,chi2016prime}.
These architectures adopt a tiled structure, as shown in \Cref{fig:overview}(a), wherein tiles are interconnected via a \ac{noc}. 
To achieve high energy efficiency and inference performance, maximizing the utilization of the \acp{pe}, located inside the tiles, is essential. This imposes a special challenge for \ac{cim} architectures since the \ac{nn}'s weights are statically assigned to the \acp{pe} and remain there during inference.
To increase the \acp{pe}'s utilization, the compiler needs to exploit both \textit{intra-} and \textit{cross-layer scheduling} of the workload~\cite{cai2023inter}. Previous research has mainly focused on intra-layer scheduling techniques \cite{yanai2016efficient,peng2019optimizing,negi2022nax,agrawal2019x}, which only consider parallel execution of individual layers.
\textit{Weight duplication}, a mapping method proposed in \cite{liu2021fpra,rhe2022vwc,zhu2018mixed},
involves assigning the same weights to multiple \acp{pe} to divide input data between them.
This approach is restricted by the limited number of \acp{pe} 
and can only accelerate individual layers.
It does not increase the utilization of the \acp{pe}.
%
%
%
%
In contrast, \textit{cross-layer scheduling} accomplishes this by considering optimizations across layer boundaries~\cite{sze2017efficient}.
Cross-layer scheduling forwards parts of a layer's \ac{ofm} to subsequent layers before the entire \ac{ofm} has been computed (\Cref{fig:overview}(d)).
While previous research has addressed the development of tiled \ac{cim} architectures that enable cross-layer scheduling in hardware~\cite{shafiee2016isaac, liu2021fpra, symons2022towards}, there is a lack of a software methodology to fully exploit this feature.
We want to close this gap by presenting the following contributions:

\begin{itemize}
\item We extend the existing weight duplication approaches by developing an algorithm that decides which parts of the \ac{nn} are duplicated to achieve minimum inference latency. Details of the TensorFlow implementation are provided.

\item We introduce CLSA-CIM, a novel approach for cross-layer scheduling on tiled \ac{cim} architectures.
CLSA-CIM seamlessly integrates with existing mapping strategies, such as weight duplication, while leveraging established intra-layer scheduling techniques.

\item In a case study, we show that CLSA-CIM has the potential to achieve up to \SI{29.2}{\times} speedup by massively improving the \ac{pe} utilization.
\end{itemize}

\section{Background and Related Work}
\label{sec:background}
\subsection{\ac{rram}-based Tiled \ac{cim} Architectures}
\label{sec:cimarchitectures}
\ac{rram} devices offer programmable conductance values and are used in crossbar structures for efficient in-memory \ac{mvm} operations~\cite{cao2021neural}.
\ac{rram} cells have a limited endurance~\cite{nail2016understanding}.
It therefore makes sense to store all \ac{nn} weights only once before inference.
This also avoids costly rewriting processes~\cite{NeuRRAM}.
Consequently, \ac{rram}-based \ac{cim} architectures typically incorporate a large number of crossbars to store the entire \ac{nn}~\cite{ankit2019puma}.
Various \ac{cim} architectures have been proposed to enable efficient and parallel \ac{mvm} execution~\cite{NeuRRAM,ankit2019puma,shafiee2016isaac,song2017pipelayer}.
Drawing from these accelerators, we will define fundamental hardware requirements that must be fulfilled to support cross-layer scheduling (see \Cref{fig:overview}(a)): 

\begin{itemize}
\item Tiles that exchange data with other tiles via a \ac{noc}.
\item All tiles operate in parallel and independently.
\item Within the tiles, there are buffers to store parts of the input and output data.
\item Due to limited buffer memory, all tiles have fast access to a global DRAM for data exchange.
\item Inside the tiles, there are crossbar(s), also called \ac{pe}(s).
\item The number of tiles and \acp{pe} is sufficient to store all weights of the \ac{nn} at least once on the architecture.
\item Each tile has \ac{gpeu} to execute other operations than \ac{mvm} (e.g., \textit{pooling}).
\end{itemize}

While the majority of the tiled \ac{cim} accelerators meet these cross-layer scheduling requirements~\cite{ankit2019puma,shafiee2016isaac,song2017pipelayer}, they differ in \ac{gpeu} logic, \ac{noc} structure, tile count, \ac{pe} dimensions, or buffer capacities.
From the cross-layer scheduling perspective, these differences are not relevant as long as the above-mentioned requirements are met.

\subsection{Intra-Layer Scheduling and Layer-by-Layer Inference}
\label{sec:layerbylayer}
The performance gains of tiled \ac{cim} accelerators stem from the fact that the \acp{pe} can efficiently perform \acp{mvm} in parallel.
Previous works exploit intra-layer parallelism to speed up inference~\cite{yanai2016efficient,peng2019optimizing,negi2022nax,agrawal2019x}. However, the overall \ac{pe} utilization remains low as only one layer's \ac{pe}(s) are active at any time, which is called layer-by-layer inference.

\subsection{Weight Duplication Mapping}
\label{sec:backgroundwdup}
To speed up layer-by-layer inference, weight duplication aims to further enhance the intra-layer parallelization capabilities of \acp{nn} by storing the same weight sets in two or more \acp{pe}. This creates parallel duplicated layer nodes in the \acp{nn} graph, which can be executed in parallel on \ac{cim} architectures (see \Cref{fig:overview}(c)).
This idea has been proposed in previous works~\cite{liu2021fpra,rhe2022vwc,zhu2018mixed}.
Weight duplication comes at the cost of increased resource requirements.
It is mandatory to determine which layers should be duplicated and how often.
We extend existing approaches by developing an algorithm that estimates the costs and values of duplicating certain weights. This is explained in more detail in \cref{sec:weightduplication}. 

\subsection{Cross-Layer Inference}
The idea of cross-layer inference is that partial results can be passed to the next \acp{pe} even before an entire layer has been fully computed. This improves the inference latency and \ac{pe} utilization.
While cross-layer inference approaches have been proposed for tiled accelerators~\cite{cai2023inter}, they have not yet been applied to \ac{cim} systems. It is particularly well-suited for \ac{cim} architectures because of their weight-stationary data flow.
Some architectures are specifically designed for this purpose. For instance, the authors of \cite{liu2021fpra} presented a \ac{cim} architecture that includes synchronization mechanisms particularly tailored to support cross-layer inference. However, they do not provide a general approach on the software side. CLSA-CIM, our software-based scheduling approach, overcomes this limitation. Implementation details are presented in \cref{sec:xinf1} and \cref{sec:xinf2}.

\section{Cross-Layer Scheduling - Preparation}
\label{sec:xinf1}

Preprocessing transforms the \ac{nn} model (from TensorFlow) into a unified structure, which serves as input for the CLSA-CIM algorithm.
It involves architecture-specific high-level optimizations, the refinement of existing mapping approaches including concepts like im2col and weight duplication, as well as the use of existing intra-layer scheduling concepts.

\subsection{High-Level Optimizations}
\textit{BN folding}: \Ac{bn} layers enhance training stability and convergence speed by normalizing the input distributions. For inference, the \ac{bn} layer can be merged with the previous operation, known as \ac{bn} folding~\cite{jacob2018quantization}.
It improves computational efficiency and memory utilization by adjusting the kernel weights $w$ and the bias $b$ in the Conv2D operation.

\textit{Partitioning}:
The \ac{nn} is divided into \textit{base layers}, i.e., operations executed on the \ac{pe} (like convolutions and dense layers), and \textit{non-base layers} (all remaining layers).
In the \ac{nn} graph, padding and bias addition are decoupled from the base layer, eliminating redundancy in the graph representation.

\begin{figure}[!bht]
  \centering
  \includegraphics[width=\linewidth, keepaspectratio]{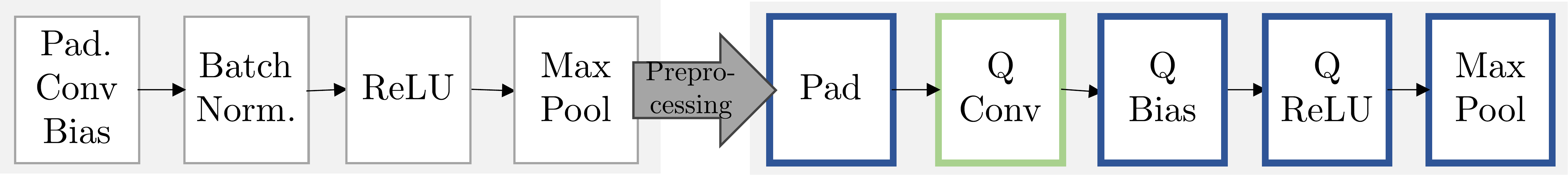}
  \caption{Partitioning, quantization (Q), and \ac{bn} folding. The resulting canonical \ac{nn} representation is split into \textit{base} (green) and \textit{non-base} (blue) layers}
  \label{fig:graphpreprocessing}
  \vspace{-0.3cm}
\end{figure}

\textit{Quantization}: Base layers need to be quantized due to the limited resolution of \ac{pe} (\ac{rram}) cells. For existing \acp{pe}, this resolution can be up to \SI{4}{Bits}~\cite{NeuRRAM}.
The preprocessing steps are summarized in \cref{fig:graphpreprocessing} using a minimal example.

\vspace{-0.3cm}

\subsection{Im2col and Intra-Layer Scheduling}
\label{sec:intralayerparallel}
\begin{figure}[thb]
  \centering
  \includegraphics[width=.9\linewidth, keepaspectratio]{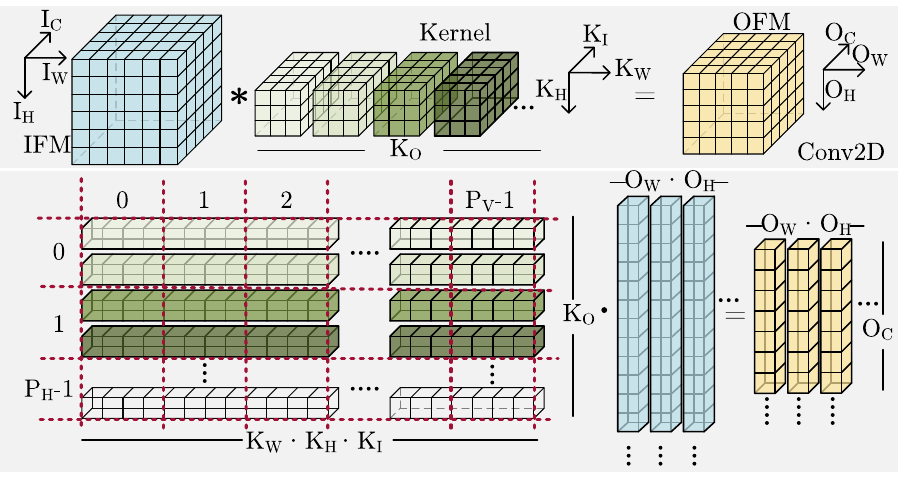}
  \caption{Conv2D to GEMM transformation using im2col}
  \label{fig:conv2dgemm}
  \vspace{-0.2cm}
\end{figure}

Base layers must be translated into \acp{mvm} to execute them on the \acp{pe}.
One widely used technique for convolutions is converting them into \acp{gemm}. This can be accomplished through the use of im2col~\cite{yanai2016efficient}.
\Cref{fig:conv2dgemm} illustrates the im2col algorithm, which unrolls the individual kernels and arranges them in columns which leads to a $(K_W\cdot K_H\cdot K_I)\times K_O$ kernel matrix.
The kernel matrix is subdivided into submatrices of size $M\times N$, which are statically mapped to the accelerators \acp{pe}~\cite{agrawal2019x}.
Previous research shows that all \acp{pe} within a layer can operate in parallel with minimal latency overhead, a concept known as intra-layer scheduling~\cite{pelke2023mapping}.

Therefore, we simplify by assuming that the calculation of an $(1\times 1\times O_C)$ \ac{ofm} vector takes place within $t_{MVM}$, which represents the \ac{mvm} latency of a \ac{pe}.
Accordingly, the total latency to compute the \ac{ofm} of a single layer using intra-layer scheduling is
$t_{OFM} = O_H \cdot O_W \cdot t_{MVM}$.

\subsection{Weight Duplication Mapping}
\label{sec:weightduplication}
Weight duplication reduces the inference latency of a single layer since the work, i.e., the input vectors, is evenly distributed among the duplicates. The latency of the Conv2D operation reduces to $t_{OFM} = \frac{1}{D} \cdot O_H \cdot O_W \cdot t_{MVM}$, where $D$ is the number of duplicates.
Duplicating the kernel matrix comes at the cost of requiring more \acp{pe} to store all weights. 
This means that weight duplication is rather beneficial for layers with high calculation latency (large $O_H \cdot O_W$ factor) and a small number of required \acp{pe}.
As discussed in \cref{sec:cimarchitectures}, it is assumed that the architecture has a sufficient number of \acp{pe} to store all weights without rewriting. If the architecture has $F$ \acp{pe} and the \ac{nn} needs $C_{num}$ \acp{pe}, with $C_{num} < F$, weight duplication can be applied to further reduce the inference latency (see \Cref{sec:backgroundwdup}).
The solution vector $\mathbf{d}$ of Optimization Problem~\ref{pro:weightdup} determines which layers should be duplicated to achieve the best inference latency:

\begin{problem}
\begin{algorithmic}
\State \textbf{minimize}: $\sum_i \frac{t_i}{d_i}$
\State \textbf{subject to}: $\mathbf{c}^T \cdot \mathbf{d} \leq F$,
\State \hspace{1.5cm} $\mathbf{d} \geq \mathbf{1}$,
\State \hspace{1.5cm} $\mathbf{d} \in \mathbb{Z_+^N}$
\end{algorithmic}
\caption{Weight Duplication}
\label{pro:weightdup}
\end{problem}

Vector $\mathbf{t}$ contains the latencies needed to calculate the \ac{ofm} of the base layers with
$\mathbf{t}^T = \left(t_{OFM_0}, t_{OFM_1}, ..., t_{OFM_{N-1}}\right)$.
Vector $\mathbf{c}$ contains the number of required $M\times N$ \acp{pe} for every base layer, e.g., convolutions (see \Cref{fig:conv2dgemm}):
\begin{align}
    c_{i} = \underbrace{\left\lceil \frac{K_{W,l} \cdot K_{H,l}  \cdot K_{I,l}}{N} \right\rceil}_{\eqqcolon P_{V,i}} \cdot
    \underbrace{\left\lceil \frac{K_{O,l}}{M} \right\rceil}_{\eqqcolon P_{H,i}}, \quad \sum_i c_i = C_{num}
\end{align}

\begin{figure}[!bht]
  \centering
  \includegraphics[width=\linewidth, keepaspectratio]{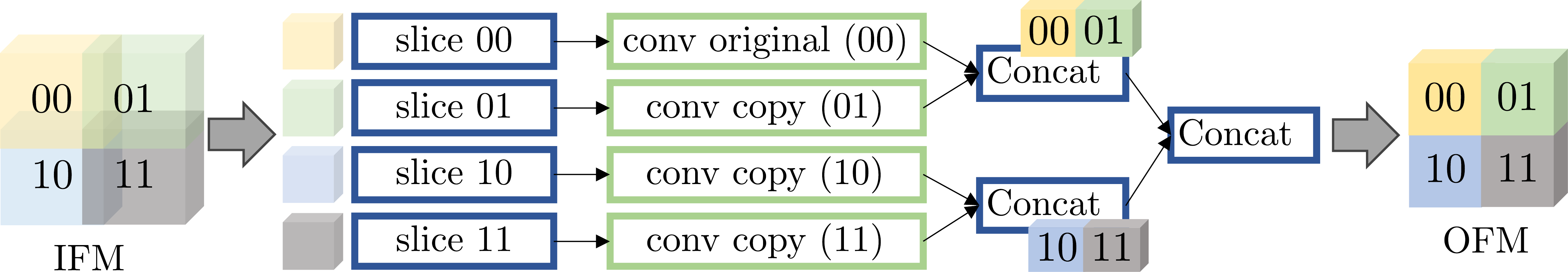}
  \caption{Implementation of weight duplication using three duplicates}
  \label{fig:weightduplication}
\end{figure}

The vector $\mathbf{d}$ also specifies the number of base layer duplicates to be created.
The latency values $\mathbf{t}_{i}$ for calculating one layer $i$ using intra-layer scheduling are set according to \Cref{sec:intralayerparallel}.
Note that the solution determines which weights to duplicate, but it does not determine how to distribute the work, i.e., the \ac{ifm}, among the duplicates.
Keeping in mind the intra-layer scheduling algorithm in \cref{sec:intralayerparallel}, the \acp{ifm} and \acp{ofm} should be cut along the $I_W$/$O_W$ and/or $I_H$/$O_H$ dimensions, as shown in \cref{fig:weightduplication}.

\vspace{-0.2cm}

The example provides details of the TensorFlow-specific implementation of weight duplication. The \ac{ofm} is divided into $2\times 2\times 1$ disjoint parts. In the \ac{nn} graph, this is realized by applying one \texttt{tf.slice} operation on the \ac{ifm} for each duplicate.
The \ac{ifm} slices may overlap depending on the kernel shape and stride.
After the distributed calculations, the \acp{ofm} are concatenated using \texttt{tf.keras.layers.Concatenate}. The depth of the concatenated tree corresponds to the number of dimensions along which it has been cut.
The influence of weight duplication on the inference latency will be discussed in \Cref{sec:casestudy}.

\section{Cross-Layer Scheduling - CLSA-CIM}
\label{sec:xinf2}
\begin{figure*}[!htbp]
  \centering
  \includegraphics[width=.95\linewidth, keepaspectratio]{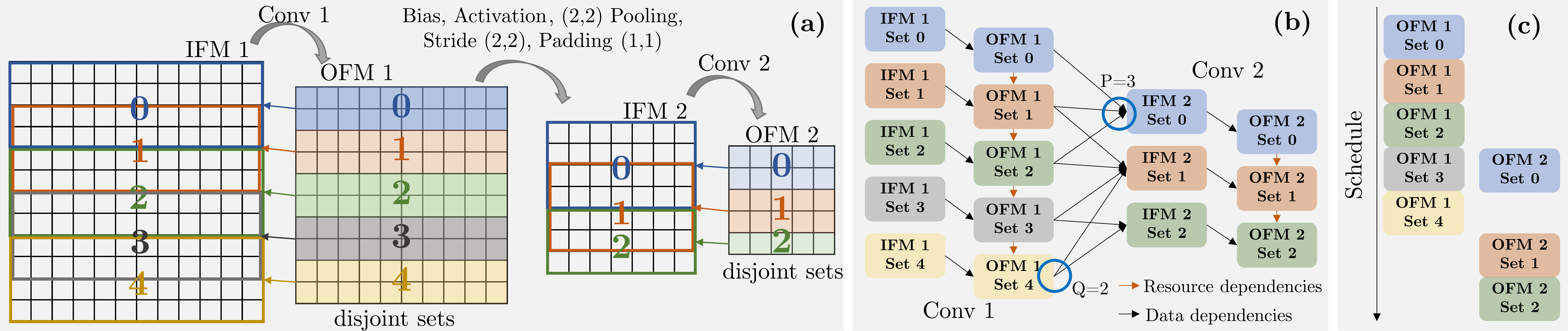}
  \caption{Minimal example for CLSA-CIM using two consecutive Conv2D layers and a non-base layer path including bias, activation, pooling, and padding: Determine sets (a), determine dependencies (b), and cross-layer scheduling (c)}
  \label{fig:algorithm}
  \vspace{-0.3cm}
\end{figure*}

CLSA-CIM builds upon the mapping and intra-layer scheduling concepts from \cref{sec:xinf1}.
It aims to minimize the inference latency by maximizing the utilization of the tiles (see \Cref{sec:cimarchitectures}).
The algorithm comprises two preprocessing stages for creating the necessary data structures, \textit{determine sets} and \textit{determine dependencies} (\cref{fig:algorithm}(a)-(b)), followed by two scheduling stages: First, intra-layer scheduling is applied, followed by the actual cross-layer scheduling (\cref{fig:algorithm}(c)).

\subsubsection{Stage I - Determine Sets}
The \acp{ofm} is divided into disjoint sets, which are the minimum scheduling units.
This means that all elements within this set must be processed before elements from another set of the same \ac{ofm} can be calculated.
The sets should ideally contain a similar number of elements; otherwise, the execution time for each set may vary.
Additionally, a hyperrectangle shape allows to identify the set's location and size using two coordinates.
Increasing the number of sets provides a more detailed scheduling granularity.
The sets should be sufficiently large to facilitate the execution of non-base layer operations, such as pooling.
In the example in \Cref{fig:algorithm}(a), the sets must contain at least $2 \times 2$ values to accommodate $(2,2)$ pooling with a stride of $(2,2)$. 
Next, the intra-layer dependencies are determined.
For each set of the \ac{ofm}, the corresponding set of the \ac{ifm} is calculated.
When adding new base layers to the algorithms, this dependency has to be specified.

\subsubsection{Stage II - Determine Dependencies}
This stage calculates the dependencies between consecutive base layers.
The two points specifying the location and size of the \ac{ofm} set of a predecessor are propagated along the non-base layer path to determine which \ac{ifm} sets are affected.
In \Cref{fig:algorithm}(b), it is evident that each \ac{ofm} set can influence multiple \ac{ifm} sets (denoted as $Q$), and likewise, each \ac{ifm} set can be affected by multiple \ac{ofm} sets (denoted as $P$).

\subsubsection{Stage III - Intra-Layer Scheduling}
In the third stage, the scheduling order of the \ac{ofm} sets is determined for each base layer individually.
The execution order of the \ac{ofm} sets can be seen in \Cref{fig:algorithm}(b). The orange-colored connections between \ac{ofm} sets of the same layer indicate resource dependencies, which means that the same crossbars are needed to calculate those sets. In the example, $OFM 1 set_0$ has to be scheduled before the other sets of $Conv1$ since it allocates the resources first.
The dependencies marked in black are data dependencies.
To generate the \ac{ifm} set of $Conv2$ from the \ac{ofm} sets of $Conv1$, non-base layer operations, e.g., pooling, are applied. 

\subsubsection{Stage IV - Cross-Layer Scheduling}
\Cref{fig:algorithm}(c) shows the resulting schedule. Note that the non-base layer operations are not illustrated due to simplicity.
CLSA-CIM ascertains the earliest feasible starting point for computing each \ac{ofm} set in the \ac{nn}. In other words, an \ac{ofm} set is scheduled once all the required \ac{ifm} sets of its predecessors have been scheduled.

\subsection{Combine Weight Duplication and Cross-Layer Scheduling}
Weight duplication is a mapping technique, whereas cross-layer inference is a scheduling technique. These concepts can be used independently.
Combining them can further reduce inference latency. The weight duplication algorithm is applied first, resulting in a non-sequential \ac{nn} graph where each layer can have multiple predecessors and successors.
CLSA-CIM is applied after that. It is designed to handle non-sequential models in a generic manner, requiring no additional modifications or adjustments. This allows for the seamless integration of weight duplication and CLSA-CIM.

\section{Evaluation}
\label{sec:results}
This chapter evaluates the performance of CLSA-CIM.
We distinguish between three approaches: weight duplication mapping combined with layer-by-layer inference (\texttt{wdup}), cross-layer inference (\texttt{xinf}), and the combination of weight duplication mapping and cross-layer inference (\texttt{wdup+xinf}).
All speedup measurements are referenced to the layer-by-layer inference (see \Cref{sec:layerbylayer}).
As there are currently no commercially available \ac{cim} chips, we use a custom system-level simulator, similar to previous works~\cite{liu2021fpra,rhe2022vwc,lu2021neurosim}. We calculate the maximum achievable utilization and minimum inference latency achievable with CLSA-CIM.
For the simulation, three core parameters are required: the number of \acp{pe}, the dimensions of a \ac{pe}, and the \ac{mvm} latency. In a case study, we assume a $256\times 256$ crossbar and an \ac{mvm} latency of $t_{MVM}=$~\SI{1400}{\ns}~\cite{NeuRRAM}, which we call a \textit{cycle}.
The number of \ac{cim} cores is kept variable in the simulation to investigate its impact on the latency.
If future \ac{cim} architectures meet the prerequisites outlined in \cref{sec:cimarchitectures}, CLSA-CIM can be used.
CLSA-CIM increases the architecture utilization $Ut$, which is defined
as the mean over the ratio of the active cycle time $t_{p, active\_cycles}$ to the total inference time of the \ac{nn} $t_{NN\_cyles}$ for \ac{pe} $p$:
\vspace{-0.2cm}
\begin{align}
\label{eq:utilization}
Ut \coloneqq \frac{1}{\#PE} \left( \sum_{p \in PE} \frac{t_{p, active\_cycles}}{t_{NN\_cyles}} \right)
\end{align}

The number of \acp{pe} is varied for each benchmark to enable weight duplication. The notation ``\texttt{wdup$_{+x}$}'', e.g., ``\texttt{wdup$_{+32}$}'', means that the architecture has $32$ \acp{pe} more than needed to store all \ac{nn} weights exactly once.

\subsection{CLSA-CIM - A Case Study}
\label{sec:casestudy}
We analyze our scheduling approach (CLSA-CIM) with a TinyYOLOv4 case study.
TinyYOLOv4 is a non-sequential \ac{nn} for object detection and classification.
\Cref{tab:tinyyolov4} shows an extract of the base layer structure of TinyYOLOv4.
\vspace{-0.1cm}
\begin{table}[!hbt]
    \centering
    \caption{\label{tab:tinyyolov4}Extract of the base layer structure of TinyYOLOv4}
    \begin{tabular}{l|cccccr}
    Layer & IFM shape & \ac{ofm} shape & \#PE & Cycles \\
     & (HWC) & (HWC) &  256$\times$256 & $t_{init}$ \\
    \hline
    conv2d     & (417, 417, 3)  & (208, 208, 32) & 1 & 43264 \\
    conv2d\_1  & (209, 209, 32) & (104, 104, 64) & 2 & 10816 \\
    conv2d\_2  & (106, 106, 64) & (104, 104, 64) & 3 & 10816 \\
    ...        & ...          & ...          &...& ... \\
    conv2d\_16 & (15, 15, 256)  & (13, 13, 512)  & 18 & 169 \\
    conv2d\_20 & (26, 26, 256)  & (26, 26, 255)  & 1 & 676 \\
    conv2d\_17 & (13, 13, 512)  & (13, 13, 255)  & 2 & 169 
    \end{tabular} 
\end{table}

\begin{figure*}[!tp]
    \centering
    \begin{subfigure}[b]{0.39\linewidth}
         \centering
    \includegraphics[width=\linewidth, keepaspectratio]{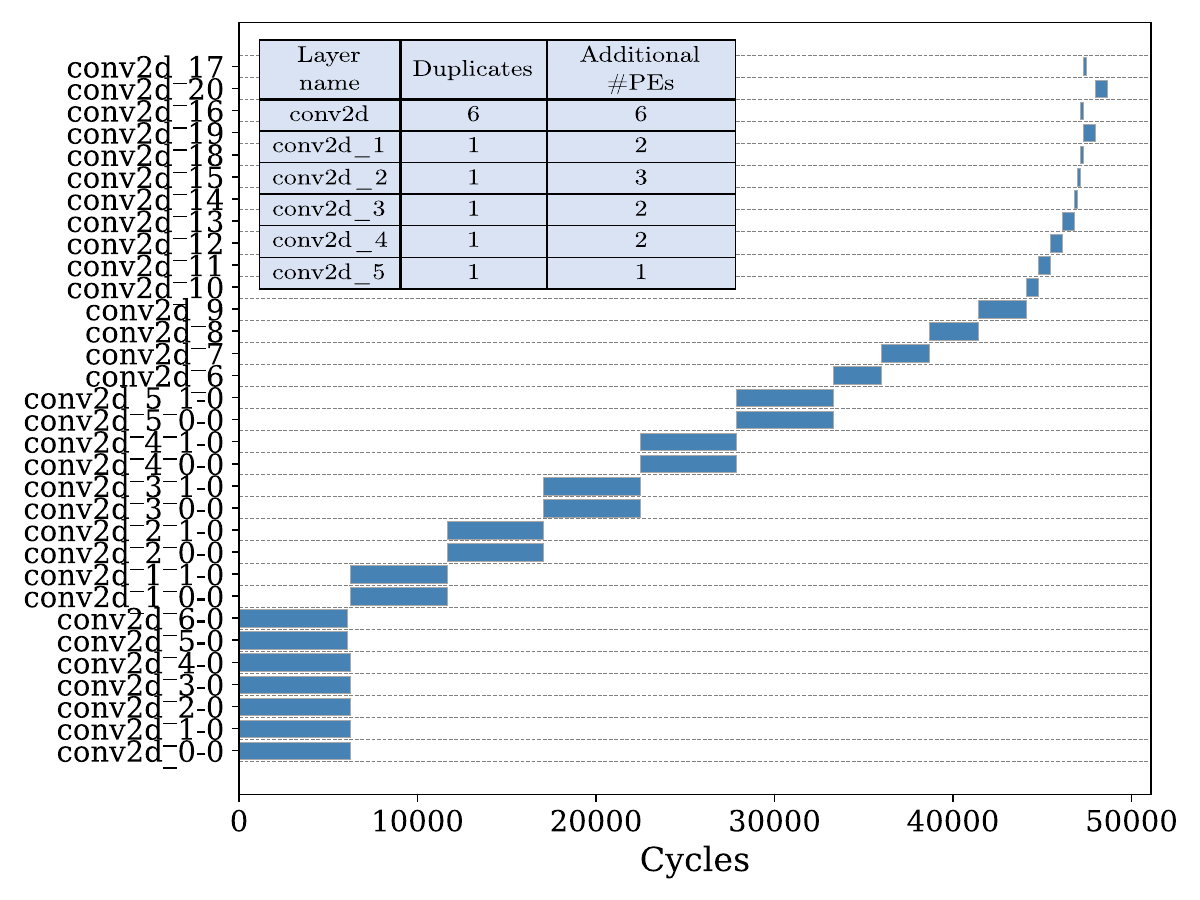}
    \caption{Weight duplication (\texttt{wdup$_{+16}$}), layer-by-layer}
    \label{fig:schedulewdup}
    \end{subfigure}
    \begin{subfigure}[b]{0.39\linewidth}
         \centering
    \includegraphics[width=\linewidth, keepaspectratio]{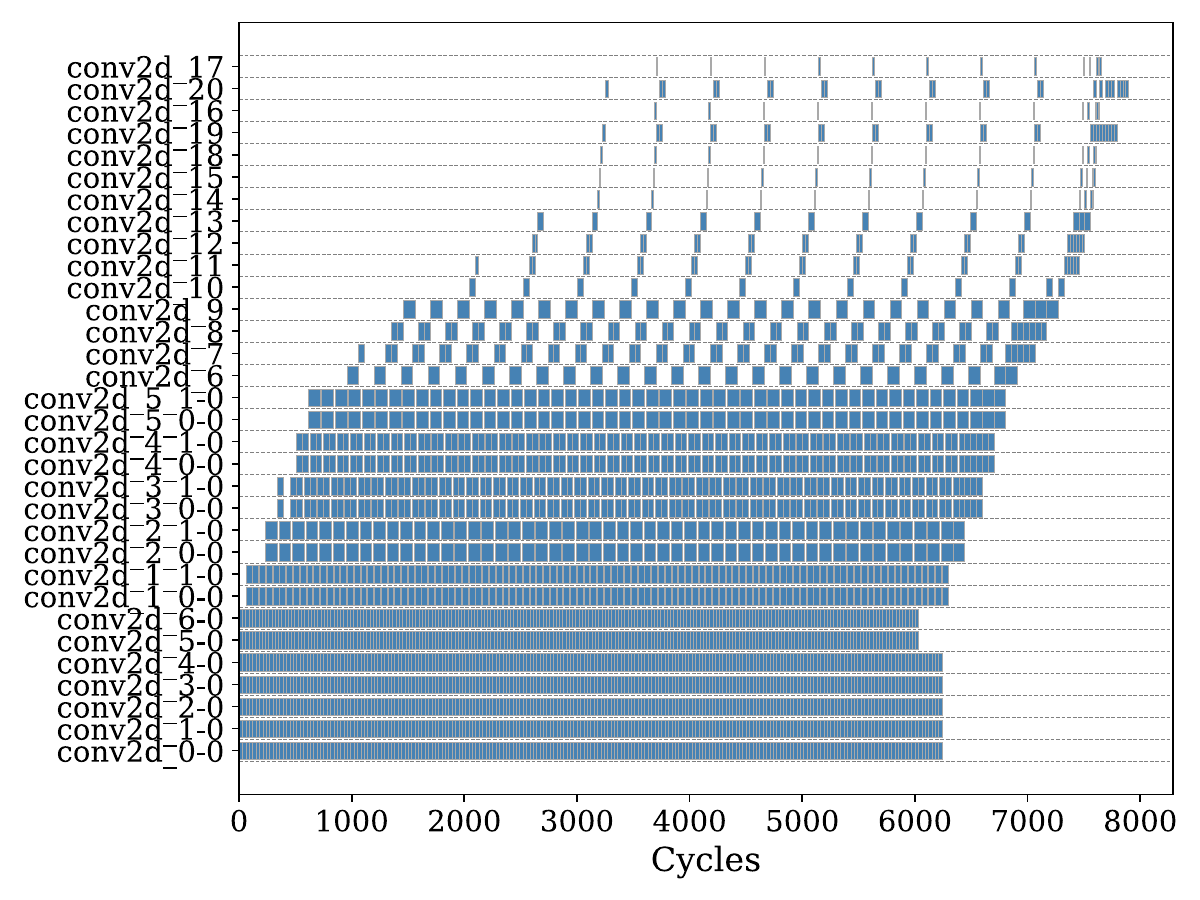}
    \caption{Weight duplication (\texttt{wdup$_{+16}$}), CLSA-CIM (\texttt{xinf})}
    \label{fig:scheduleweightdupcrosslayer}
    \end{subfigure}
    \begin{subfigure}[b]{0.19\linewidth}
         \centering
    \includegraphics[width=\linewidth, keepaspectratio]{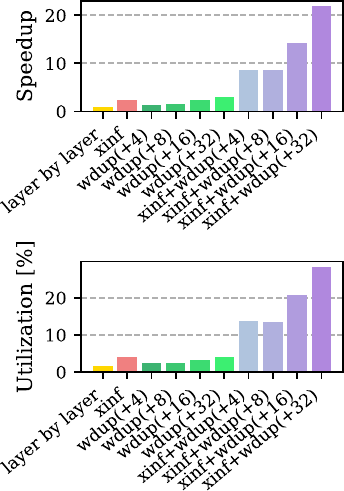}
    \caption{Speedup and utilization}
    \label{fig:tinyyolov4_lat_ut}
    \end{subfigure}
    \caption{Visualization of weight duplication mapping (a) and CLSA-CIM (b) using $x=16$ additional \acp{pe}, speedup and \ac{pe} utilization for different weight mapping (weight duplication) and scheduling (layer-by-layer, CLSA-CIM) combinations (c)
        \label{fig:tinyyolocasestudy}\vspace{-0.2cm}}
\end{figure*}

TinyYOLOv4 has $18$ Conv2D layers.
The minimum number of \acp{pe} required to store all weights at least once, $PE_{min}$, is $117$.
The time $t_{init}$ is the duration of executing the layer (in cycles) using only intra-layer scheduling (see \cref{sec:intralayerparallel}).
Since the $O_H \cdot O_W$ factor is higher for the first layers, they are more time-consuming to compute. The first layers need fewer \acp{pe}, which makes them a good choice for weight duplication.

\Cref{fig:tinyyolocasestudy} provides an illustration of the weight duplication mapping (\texttt{wdup}) of the TinyYOLOv4 benchmark combined with layer-by-layer scheduling (\cref{fig:schedulewdup}) and CLSA-CIM (\cref{fig:scheduleweightdupcrosslayer}).
The solution of \cref{pro:weightdup} in \cref{sec:weightduplication} reveals that for $x=16$ additional \acp{pe}, the first $6$ Conv2D layers need to be duplicated according to the table in \cref{fig:schedulewdup}.
\cref{fig:tinyyolov4_lat_ut} confirms what is visible in \cref{fig:scheduleweightdupcrosslayer}: CLSA-CIM (\texttt{xinf}) increases the utilization of the \acp{pe} to a total of \SI{4.1}{\%}.
In combination with weight duplication and $x=32$ additional \acp{pe} (\texttt{wdup$_{+32}$}), in total $117+32$ \acp{pe}, the utilization increases up to \SI{28.4}{\%}.
This corresponds to an inference speedup of up to \SI{21.9}{\times}.
The relationship between speedup $S$ and utilization $Ut$ for \texttt{+x} \acp{pe} and configuration $c$ is 

\vspace{-0.3cm}
\begin{align}
\label{eq:speedupVSutilization}
S_{x, c} \approx \frac{Ut_{x, c} \cdot (PE_{min}+x)}{Ut_{layer\_by\_layer} \cdot PE_{min}}.
\end{align}

\subsection{Performance Results}


We further evaluate benchmarks that have a higher demand for \acp{pe} than TinyYOLOv4.
This includes sequential models like VGG16 and VGG19, and non-sequential models like TinyYOLOv3, ResNet50, ResNet101, and ResNet152.
Since the latency and utilization depend on the dimensions of the \ac{ifm}, the dimensions are listed in \Cref{tab:benchmarkInputShape}. 
\begin{table}[!hbt]
    \centering
    \caption{\label{tab:benchmarkInputShape}List of benchmarks}
    \begin{tabular}{l|ccc}
    Benchmark & Input shape & Base layers & Min. \# required\\
     & (HWC) & (number) & 256$\times$256 PEs\\
    \hline
    TinyYOLOv3 & (416, 416, 3) & 13 & 142\\
    VGG16 & (224, 224, 3) & 13 & 233\\
    VGG19 & (224, 224, 3) & 16 & 314\\
    ResNet50 & (224, 224, 3) & 53 & 390\\
    ResNet101 & (224, 224, 3) & 104 & 679\\
    ResNet152 & (224, 224, 3) & 155 & 936\\
    \end{tabular} 
\end{table} 

\begin{figure*}[t!]
    \centering
    \begin{subfigure}[b]{0.49\linewidth}
         \centering
    \includegraphics[width=\linewidth, keepaspectratio]{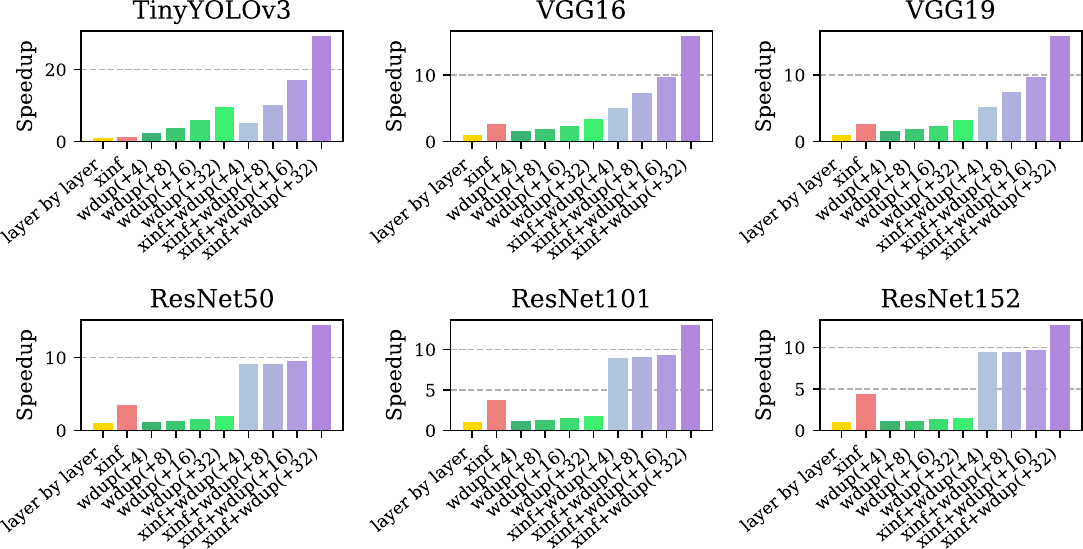}
    \caption{Inference speedup in relation to layer-by-layer scheduling}
    \label{fig:multinnlatency}
    \end{subfigure}
    \begin{subfigure}[b]{0.49\linewidth}
         \centering
    \includegraphics[width=\linewidth, keepaspectratio]{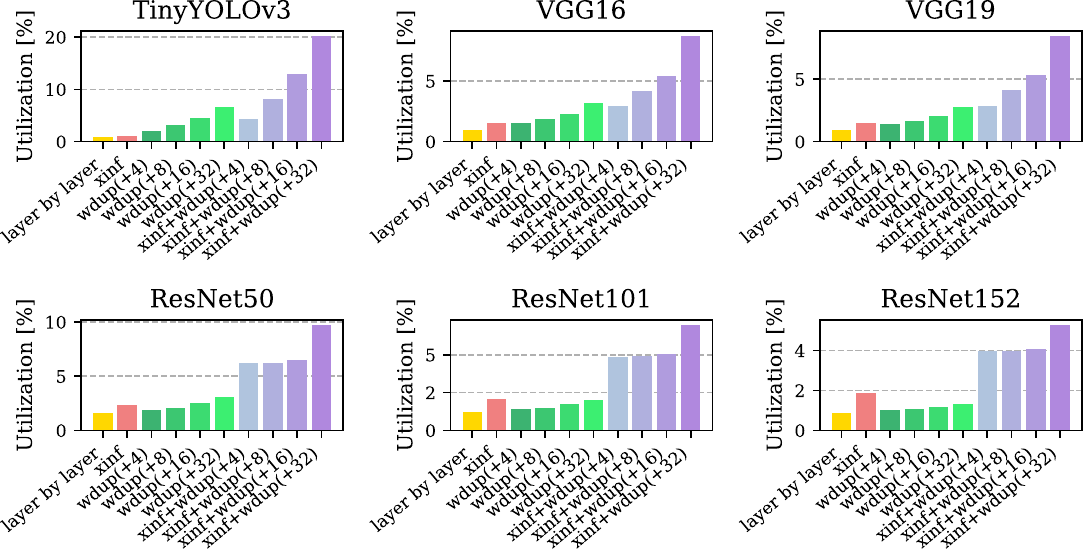}
    \caption{Utilization in relation to layer-by-layer scheduling}
    \label{fig:multinnutilization}
    \end{subfigure}
    \caption{Combinations of mapping and scheduling in contrast to layer-by-layer scheduling without weight duplication: layer-by-layer scheduling with weight duplication (\texttt{wdup}), CLSA-CIM (\texttt{xinf}), and weight duplication with CLSA-CIM (\texttt{wdup+xinf})\label{fig:multinnresults}\vspace{-0.2cm}}
\vspace{-0.3cm}
\end{figure*}

The inference latency speedups and \ac{pe} utilizations are compared for different combinations of mapping (\texttt{wdup$_{+x}$}) and scheduling (\texttt{layer-by-layer}, \texttt{xinf}).
For additional \acp{pe}, we consider the setups $x \in \{4,8,16,32\}$, i.e., for VGG19, $314$ to $346$ PEs. This enables a better comparison across different benchmarks.
In the tested configurations in \cref{fig:multinnlatency}, pure weight duplication yields a modest speedup for large models, from \SI{1.1}{\times} to \SI{1.9}{\times}.
This is because the number of additional \acp{pe} (up to $32$) is small compared to the minimum required \acp{pe} to store the entire \ac{nn} on the accelerator.
 CLSA-CIM (\texttt{xinf}) achieves a speedup of up to \SI{4.4}{\times} for large models compared to layer-by-layer scheduling.
The best results are achieved by combining CLSA-CIM and weight duplication. This approach yields the highest speedup of \SI{29.2}{\times} for the TinyYOLOv3 benchmark.
Of particular interest is that only $x=4$ additional \acp{pe} are sufficient to outperform the pure \texttt{xinf} configuration by a factor of almost \SI{2}{\times}.
We observe this even for ResNet152, where $x=4$ \acp{pe} is very small compared to the minimum \ac{pe} requirement of $936$.
This can be attributed to the fact that, as demonstrated in \cref{fig:schedulewdup}, the first layer is relatively computation-intensive.
%
%
%
%
\Cref{fig:multinnutilization} illustrates the \ac{pe} utilizations for each benchmark.
The utilization is increased by CLSA-CIM across all benchmarks, surpassing the impact of pure weight duplication.
Again, the combination of weight duplication and CLSA-CIM delivers the best performance values.
For smaller models, higher utilization rates can be achieved, with TinyYOLOv3 reaching a maximum utilization of $20.1\%$. This represents an improvement of \SI{17.9}{\times} compared to layer-by-layer scheduling.
Since the final layers often require many \acp{pe} (see \cref{tab:tinyyolov4}), but at the same time are less computationally intensive, the utilization of the architecture for a single \ac{nn} inference usually remains below \SI{10}{\percent}.
As the model depth increases, the utilization decreases, as observed in the ResNet benchmarks. This is due to the limited parallelization capabilities between layers which are far apart in the \ac{nn} graph.

\subsection{Limitations and Future Work}
As mentioned in \Cref{sec:cimarchitectures}, our current work focuses on cases where the number of crossbars is sufficient to accommodate complete \acp{nn} on the architecture. However, in future research, we aim to explore more general scenarios. CLSA-CIM is already designed to accept the crossbar dimensions as an input parameter, allowing for adaptability to arbitrary sizes.
It is important to acknowledge that the speedup values presented in this study represent peak performance. There may be architecture-dependent factors that could potentially impact latency. For example, the costs associated with data movement have not been differentiated yet. Depending on the topology, forwarding partial results may incur varying costs. Furthermore, it is possible for cores to share resources such as adders, further imposing constraints on scheduling algorithms.
Our future work will involve extending our abstract architecture description to account for these factors, enabling full architecture retargetability.

\section{Conclusion}

The development of efficient scheduling algorithms for tiled \ac{cim} architectures is crucial to fully utilize the potential of \ac{cim} concepts.
Our scheduling approach, CLSA-CIM, enables cross-layer inference on top of existing intra-layer scheduling and weight duplication mapping algorithms, which significantly enhances the utilization of \acp{pe} of up to \SI{17.9}{\times}, resulting in an inference speedup of up to \SI{29.2}{\times}.
We conducted evaluations using state-of-the-art \acp{nn}, including a case study of the TinyYOLOv4 model to visualize the algorithms.
In summary, our work contributes to the advancement of scheduling approaches and algorithms for \ac{cim} architectures. It sheds light on the benefits of combining cross-layer inference and weight duplication, paving the way for enhanced performance of \ac{ml} applications on \ac{cim} architectures.

\bibliographystyle{IEEEtran}
\bibliography{bibtexentry}

\begin{thebibliography}{10}
\providecommand{\url}[1]{#1}
\csname url@samestyle\endcsname
\providecommand{\newblock}{\relax}
\providecommand{\bibinfo}[2]{#2}
\providecommand{\BIBentrySTDinterwordspacing}{\spaceskip=0pt\relax}
\providecommand{\BIBentryALTinterwordstretchfactor}{4}
\providecommand{\BIBentryALTinterwordspacing}{\spaceskip=\fontdimen2\font plus
\BIBentryALTinterwordstretchfactor\fontdimen3\font minus
  \fontdimen4\font\relax}
\providecommand{\BIBforeignlanguage}[2]{{%
\expandafter\ifx\csname l@#1\endcsname\relax
\typeout{** WARNING: IEEEtran.bst: No hyphenation pattern has been}%
\typeout{** loaded for the language `#1'. Using the pattern for}%
\typeout{** the default language instead.}%
\else
\language=\csname l@#1\endcsname
\fi
#2}}
\providecommand{\BIBdecl}{\relax}
\BIBdecl

\bibitem{zou2021breaking}
X.~Zou, S.~Xu, X.~Chen, L.~Yan, and Y.~Han, ``{Breaking the von Neumann
  bottleneck: architecture-level processing-in-memory technology},''
  \emph{Science China Information Sciences}, 2021.

\bibitem{chang2017memcomputing}
Y.-F. Chang \emph{et~al.}, ``{Memcomputing (Memristor + Computing) in Intrinsic
  SiOx-Based Resistive Switching Memory: Arithmetic Operations for Logic
  Applications},'' \emph{IEEE (T-ED)}, 2017.

\bibitem{vetter2015opportunities}
J.~S. Vetter and S.~Mittal, ``{Opportunities for Nonvolatile Memory Systems in
  Extreme-Scale High Performance Computing},'' \emph{Computing in Science \&
  Engineering}, 2015.

\bibitem{NeuRRAM}
W.~Wan \emph{et~al.}, ``{A compute-in-memory chip based on resistive
  random-access memory},'' \emph{Nature}, 2022.

\bibitem{ankit2019puma}
A.~Ankit \emph{et~al.}, ``{PUMA: A Programmable Ultra-efficient Memristor-based
  Accelerator for Machine Learning Inference},'' in \emph{ASPLOS XXIV}, 2019.

\bibitem{shafiee2016isaac}
A.~Shafiee \emph{et~al.}, ``{ISAAC: A convolutional neural network accelerator
  with in-situ analog arithmetic in crossbars},'' \emph{ACM SIGARCH Computer
  Architecture News}, 2016.

\bibitem{chi2016prime}
P.~Chi \emph{et~al.}, ``{Prime: A novel processing-in-memory architecture for
  neural network computation in reram-based main memory},'' \emph{ACM SIGARCH
  Computer Architecture News}, 2016.

\bibitem{cai2023inter}
J.~Cai, Y.~Wei, Z.~Wu, S.~Peng, and K.~Ma, ``{Inter-layer Scheduling Space
  Definition and Exploration for Tiled Accelerators},'' in \emph{50th ISCA},
  2023.

\bibitem{yanai2016efficient}
K.~Yanai, R.~Tanno, and K.~Okamoto, ``{Efficient Mobile Implementation of A
  CNN-based Object Recognition System},'' in \emph{Proceedings of the 24th ACM
  international conference on Multimedia}, 2016.

\bibitem{peng2019optimizing}
X.~Peng, R.~Liu, and S.~Yu, ``{Optimizing Weight Mapping and Data Flow for
  Convolutional Neural Networks on Processing-in-Memory Architectures},''
  \emph{IEEE Transactions on Circuits and Systems Is}, 2019.

\bibitem{negi2022nax}
S.~Negi, I.~Chakraborty, A.~Ankit, and K.~Roy, ``{NAX: neural architecture and
  memristive xbar based accelerator co-design},'' in \emph{DAC}, 2022.

\bibitem{agrawal2019x}
A.~Agrawal, C.~Lee, and K.~Roy, ``{X-CHANGR: Changing Memristive Crossbar
  Mapping for Mitigating Line-Resistance Induced Accuracy Degradation in Deep
  Neural Networkss},'' \emph{arXiv preprint arXiv:1907.00285}, 2019.

\bibitem{liu2021fpra}
X.~Liu \emph{et~al.}, ``{FPRA: A Fine-grained Parallel RRAM Architecture},'' in
  \emph{2021 IEEE/ACM ISLPED}.\hskip 1em plus 0.5em minus 0.4em\relax IEEE,
  2021.

\bibitem{rhe2022vwc}
J.~Rhe, S.~Moon, and J.~H. Ko, ``{VWC-SDK: Convolutional Weight Mapping Using
  Shifted and Duplicated Kernel with Variable Windows and Channels},''
  \emph{IEEE JETCAS}, 2022.

\bibitem{zhu2018mixed}
Z.~Zhu \emph{et~al.}, ``{Mixed Size Crossbar based RRAM CNN Accelerator with
  Overlapped Mapping Method},'' in \emph{IEEE/ACM ICCAD}, 2018.

\bibitem{sze2017efficient}
V.~Sze, Y.-H. Chen, T.-J. Yang, and J.~S. Emer, ``{Efficient Processing of Deep
  Neural Networks: A Tutorial and Survey},'' \emph{Proceedings of the IEEE},
  2017.

\bibitem{symons2022towards}
A.~Symons, L.~Mei, S.~Colleman, P.~Houshmand, S.~Karl, and M.~Verhelst,
  ``{Towards Heterogeneous Multi-core Accelerators Exploiting Fine-grained
  Scheduling of Layer-Fused Deep Neural Networks},'' \emph{arXiv preprint
  arXiv:2212.10612}, 2022.

\bibitem{cao2021neural}
W.~Cao, Y.~Zhao, A.~Boloor, Y.~Han, X.~Zhang, and L.~Jiang, ``{Neural-PIM:
  Efficient Processing-In-Memory With Neural Approximation of Peripherals},''
  \emph{IEEE Transactions on Computers}, 2021.

\bibitem{nail2016understanding}
C.~Nail \emph{et~al.}, ``Understanding rram endurance, retention and window
  margin trade-off using experimental results and simulations,'' in \emph{IEEE
  IEDM}.\hskip 1em plus 0.5em minus 0.4em\relax IEEE, 2016.

\bibitem{song2017pipelayer}
L.~Song, X.~Qian, H.~Li, and Y.~Chen, ``{PipeLayer: A Pipelined ReRAM-Based
  Accelerator for Deep Learning},'' in \emph{IEEE HPCA}, 2017.

\bibitem{jacob2018quantization}
B.~Jacob \emph{et~al.}, ``{Quantization and Training of Neural Networks for
  Efficient Integer-Arithmetic-Only Inference},'' in \emph{CVPR}, 2018.

\bibitem{pelke2023mapping}
R.~Pelke, N.~Bosbach, J.~Cubero, F.~Staudigl, R.~Leupers, and J.~M. Joseph,
  ``{Mapping of CNNs on multi-core RRAM-based CIM architectures},'' in
  \emph{IFIP/IEEE VLSI-SoC}.\hskip 1em plus 0.5em minus 0.4em\relax IEEE, 2023.

\bibitem{lu2021neurosim}
A.~Lu, X.~Peng, W.~Li, H.~Jiang, and S.~Yu, ``{NeuroSim Simulator for
  Compute-in-Memory Hardware Accelerator: Validation and Benchmark},''
  \emph{Frontiers in artificial intelligence}, 2021.

\end{thebibliography}

\end{document}